\documentclass[pre,amsmath,amssymb,twocolumn,floatfix,showpacs]{revtex4}

\usepackage{graphicx}
\usepackage{epsfig}
\usepackage{times}
\usepackage{amsmath}
\usepackage{amssymb}
\usepackage{upgreek}

\begin{document}
\date{\today}
\title{Hybrid simulations of lateral diffusion in fluctuating
  membranes} 
\author{Ellen Reister-Gottfried}
\email{reister@theo2.physik.uni-stuttgart.de} 

\author{Stefan M.~Leitenberger}
\author{Udo Seifert}
\affiliation{{II}.~Institut f\"ur Theoretische 
  Physik, Universit\"at Stuttgart,  70550 Stuttgart, Germany}
                                %
\begin{abstract}
  In this paper we introduce a novel method to simulate lateral diffusion of
  inclusions in a fluctuating membrane. The regarded systems are governed by
  two dynamic processes: the height fluctuations of the membrane and the
  diffusion of the inclusion along the membrane. While membrane fluctuations
  can be expressed in terms of a dynamic equation which follows from the
  Helfrich Hamiltonian, the dynamics of the diffusing particle is described by
  a Langevin or Smoluchowski equation. In the latter equations, the curvature
  of the surface needs to be accounted for, which makes particle diffusion a
  function of membrane fluctuations. In our scheme these coupled dynamic
  equations, the membrane equation and the Langevin equation for the particle,
  are numerically integrated to simulate diffusion in a membrane. The
  simulations are used to study the ratio of the diffusion coefficient
  projected on a flat plane and the intramembrane diffusion coefficient for
  the case of free diffusion. We compare our results with recent analytical
  results that employ a preaveraging approximation and analyze the validity of
  this approximation. A detailed simulation study of the relevant correlation
  functions reveals a surprisingly large range where the approximation is
  applicable.
\end{abstract}
\pacs{87.15.Vv, 87.16.Dg, 87.16.Ac, 05.40.-a}
\maketitle
\section{Introduction}
During the last decade it has become apparent that lateral diffusion of
various components of the cell along the membrane is crucial for several
cellular processes, like exo- and endocytosis, cell signaling, or cell
movement.  To study all these different aspects of lateral diffusion in more
detail a whole variety of experimental methods has been developed, including
fluorescence recovery after photobleaching
(FRAP)~\cite{Lippincott:2003,Reits:2001}, single particle tracking
(SPT)~\cite{Kusumi:2005}, fluorescence correlation spectroscopy
(FCS)~\cite{Kohl:2005}, or pulsed field gradient nuclear magnetic resonance
(pfg-NMR)~\cite{Oraedd:2005a}. While the accuracy of measured diffusion
coefficients achieved with these experimental methods can be very
high~\cite{Lommerse:2004}, the interpretation of the results is often very
difficult. It is, therefore, likely that theoretical calculations and
simulations in particular will play a key role in developing a better
understanding of diffusive processes in biological membranes.

In order to correctly interpret experimental results it is necessary to
analyze what information of the system in which diffusion takes place is
documented but also what is neglected or insufficiently regarded during the
measurement. Lateral diffusion of proteins in a membrane must not be viewed as
diffusion on a flat surface: due to the flexibility of biological membranes
lateral diffusion always takes place on curved surfaces, whereby the shape of
this surface may also be time dependent. This is an important aspect that
needs to be taken into account in the experimental data analysis.  However,
this turns out to be rather difficult, because not always information on the
membrane shape can be acquired. So far several theoretical studies of free
diffusion on temporally fixed curved surfaces have been undertaken. One of the
first studies of this kind was performed by Aizenbud and
Gershon~\cite{Aizenbud:1982} who numerically solved the Smoluchowski equation
of free diffusion on a periodic surface. In later studies that include both
sophisticated numerical and analytical calculations diffusion on more
complicated surfaces is
regarded~\cite{Holyst:1999,Plewczynski:2000,Balakrishnan:2000,Faraudo:2002,Christensen:2004}.
Recently Schwartz et al.~\cite{Schwartz:2005} and Sbalzarini et
al.~\cite{Sbalzarini:2006} simulated free diffusion on surfaces reconstructed
from experimental image data. In the latter work, that analyses diffusion in
the endoplasmic reticulum (ER), experimental FRAP curves are compared with
simulation results. Good agreement is found, however, the authors point out
that the evaluation of the FRAP curves assuming the membrane to be flat leads
to diffusion coefficients that differ by a factor of about two from the actual
intramembrane diffusion coefficient. Furthermore, the experimentally
determined diffusion coefficient appears anisotropic while real diffusion
along the ER is purely isotropic.

These studies clearly show that the curvature of membranes may not be
neglected during the evaluation of experimental data.  But even if the shape
of the membrane within the measurement volume appears flat on average, the
actual shape fluctuates due to thermal activation.  These membrane
fluctuations have been thoroughly studied. While the fluctuation spectrum of a
free and almost flat membrane is easily calculated~\cite{Brochard,
  Seifert:1997}, the influence of various geometric confinements of the
membrane have also been regarded.  These include the attachment of the
membrane to a reference plane via a regular mesh of harmonic
springs~\cite{Lin:2004,Lin:2004a,Merath:2005} which resembles a model for the
attachment of the cell membrane to the cytoskeleton, a membrane that is close
to a flat non-impenetrable surface~\cite{Gov:2003}, or the inclusion of active
proteins in the membrane, that exert an out-of-plane force on the
membrane~\cite{Lin:2006}. Lin and Brown introduced a very powerful method to
simulate membrane fluctuations~\cite{Lin:2004,Lin:2004a,Lin:2005}, the Fourier
space Brownian dynamics algorithm, that allows to add a whole variety of
external influences on the membrane. Part of the simulation algorithm
introduced in this paper is closely related to this method.

Returning to the movement of proteins along the membrane, it is obvious that
membrane fluctuations will influence the measured value of diffusion
coefficients. This was first pointed out by Gustafsson and
Halle~\cite{Halle:1997,Gustafsson:1997}. But, although their work is almost
ten years old, experimentally diffusion coefficients are still determined by
use of the projected flat path particles perform instead of the real path
along the membrane. Only recently the quantitative influence of neglecting
thermal membrane fluctuations on the measured diffusion coefficients has been
estimated both by us and Gov~\cite{Reister:2005,Gov:2006}. However, these
calculations make use of a preaveraging approximation, that a priori implies
that the time it takes a diffusing particle on average to diffuse the length
$\upxi$ should be much smaller than the correlation time of a membrane
fluctuation mode with wave number $2\pi/\upxi$.  In this paper we introduce a
new algorithm that does not depend on this kind of approximation, because the
diffusive motion of a protein along a fluctuating membrane is simulated
explicitly. Clearly, the system dynamics is described by two coupled
processes: the fluctuations of the membrane and the diffusion of the particle.
The simulation of membrane fluctuations is effectively regarded by numerically
integrating the equation of motion for a membrane given in the Monge gauge,
discretely in time. At each discrete timestep we also update the position of
the diffusing particle.  To this end the discrete version of the Langevin
equation valid for the particle movement is used. Special care needs to be
taken regarding this Langevin equation because the movement of the particle
depends on the actual shape of the membrane.

As a first test we analyze the height-height correlation function of membrane
fluctuations, and compare the simulation results with the analytical result.
As mentioned above we have previously calculated the ratio of the measured, or
projected, and the intramembrane diffusion coefficient as a function of
membrane parameters within a preaveraging approximation.

Using our simulation scheme that does not rely on this kind of approximation
we determine the same ratio of diffusion coefficients by analyzing the mean
square displacement of diffusing particles. We compare the simulation results
with the analytical results and finally discuss the applicability of the
preaveraging approximation for situations when diffusive timescales are
comparable or smaller than typical membrane time scales by analyzing relevant
correlation functions.

The paper is organized as follows: While we introduce the dynamics of
membranes in the next section, we explain diffusion on a curved surface in
sec.~\ref{sec:diff_curve}. Generally diffusion can either be expressed by a
Fokker-Planck equation that gives the dynamics of the probability of finding a
particle at a certain time and position, or a Langevin equation that
corresponds to the equation of motion of the particle position. In
secs.~\ref{sec:smolouchovsky} and~\ref{sec:langevin} the Smoluchowski
equation, a particular form of the Fokker-Planck equation, and the Langevin
equation for diffusion on a curved surface expressed in the Monge gauge are
described, respectively. Since simulation results are compared with previous
calculations we briefly explain these in sec.~\ref{sec:analytical}. After
establishing the theoretical foundations necessary for this study we describe
our simulation scheme in sec.~\ref{sec:simulation} with which the results in
sec.~\ref{sec:results} are achieved. In the result section we first analyze
pure membrane fluctuations and then determine the ratio of projected and
intramembrane diffusion as a function of the membrane parameters bending
rigidity and effective surface and as a function of the intramembrane
diffusion coefficient.  After the comparison of simulation results with
analytical calculations we estimate the limits of the preaveraging
approximation in sec.~\ref{sec:sim_corr}.  The paper finishes with some
conclusions and an outlook for future work.
\section{Dynamical membrane fluctuations}
The shape of a membrane without spontaneous curvature, i.e., a membrane that is
on average flat and without overhangs, is conveniently described in the Monge
gauge. Hereby a position $\mathbf{r}^{\star}$ of the membrane is given by
$\mathbf{r}^{\star}\equiv(x,y,h(x,y))$. In the following the vector $\mathbf{r}\equiv(x,y)$
denotes the projected position in the $(x,y)$-plane. The height function
$h(\mathbf{r})$ is the distance between the membrane and the flat
$(x,y)$-plane. The energy of a membrane with bending rigidity $\kappa$
and effective surface tension $\sigma$ is given by
\begin{equation}
  \begin{split}
    \mathcal{H}&[h(\mathbf{r})]=\\&\int_Ad^2\mathbf{r}\left\{\frac{\kappa}{2}(\nabla^2h(\mathbf{r}))^2+\frac{\sigma}{2}(\nabla
      h(\mathbf{r}))^2\right\}+\mathcal{H}_1[h(\mathbf{r}),\mathbf{R}]
  \end{split}
  \label{eq:Helfrich}
\end{equation}
in the Monge gauge~\cite{Canham,Helfrich,Seifert:1997,Safran}. The projected
area of the membrane is given by $A$. A possible energy contribution
$\mathcal{H}_1[h(\mathbf{r}),\mathbf{R}]$ is caused by the inserted protein at
position $\mathbf{R}\equiv(X,Y)$ and may also depend on the membrane shape. In
the following we assume that the system's energy does not depend on the
position of the particle and will therefore drop this additional energy term.
The dynamics of the membrane is given by the equation of motion for the height
function $h(\mathbf{r})$. Because $h(\mathbf{r})$ is not a conserved order
parameter the following dynamics
applies~\cite{Seifert:1997,Lin:2004,Lin:2005}:
\begin{equation}
  \begin{split}
    \frac{\partial h(\mathbf{r},t)}{\partial t}=&-\int_Ad^2\mathbf{r}'\left\{
      \Lambda(\mathbf{r}-\mathbf{r}')\frac{\delta\mathcal{H}[h(\mathbf{r}',t)]}{\delta
        h(\mathbf{r}',t)}\right\}+\xi(\mathbf{r},t)\\
    =\int_A\!\!d^2\mathbf{r}'&\left\{\Lambda(\mathbf{r}\!-\!\mathbf{r}')\left[\kappa\nabla^4h(\mathbf{r}',t)\!-\!\sigma\nabla^2h(\mathbf{r}',t)\right]\right\}\!+\!\xi(\mathbf{r},t),
  \end{split}
  \label{eq:eom_r}
\end{equation}
where $\Lambda(\mathbf{r}-\mathbf{r}')$ is an Onsager coefficient and
$\xi(\mathbf{r},t)$ is a fluctuating force that obeys the
fluctuation-dissipation theorem:
\begin{eqnarray}
  \langle\xi(\mathbf{r},t)\rangle&=&0\\
  \langle\xi(\mathbf{r},t)\xi(\mathbf{r}',t')\rangle&=&2k_BT\,A\,\Lambda(\mathbf{r}-\mathbf{r}')\delta(t-t').
\end{eqnarray}
Applying the following Fourier transform
\begin{eqnarray}
  h(\mathbf{k},t)&=&\int_A\!d^2\mathbf{r}\,h(\mathbf{r},t)e^{-i\mathbf{k}\cdot\mathbf{r}}\\
  h(\mathbf{r},t)&=&\frac{1}{A}\sum_{\mathbf{k}}h(\mathbf{k},t)e^{i\mathbf{k}\cdot\mathbf{r}},
\end{eqnarray}
the equation of motion for the membrane becomes:
\begin{equation}
  \frac{\partial h(\mathbf{k},t)}{\partial t}=-\Lambda(\mathbf{k})\left[\kappa
    k^4+\sigma k^2\right]h(\mathbf{k},t)+\xi(\mathbf{k},t),
  \label{eq:eom_k}
\end{equation}
with
\begin{eqnarray}
  \langle\xi(\mathbf{k},t)\rangle&=&0\\
  \langle\xi(\mathbf{k},t)\xi(\mathbf{k}',t')\rangle&=&2k_BT\,A\,\Lambda(\mathbf{k})\delta_{\mathbf{k},-\mathbf{k}'}\delta(t-t').\label{eq:fluct_diss_k}
\end{eqnarray}
Both the height function $h(\mathbf{r},t)$ and the random force
$\xi(\mathbf{r},t)$ are real quantities. Therefore the relation
$h^*(\mathbf{k},t)=h(-\mathbf{k},t)$ applies (for $\xi$ respectively); 
the asterisk resembles the complex conjugate.

The Onsager coefficient for an almost planar membrane can be derived from the
Oseen tensor and takes the following form in Fourier space~\cite{Doi, Brochard, Seifert:1997}:
\begin{equation}
  \Lambda(\mathbf{k})=\frac{1}{4\eta k},
  \label{eq:Onsager}
\end{equation}
with viscosity $\eta$ of the fluid surrounding the membrane. 

Due to the linearity of eq.~\eqref{eq:eom_k} the height-height correlation
function $\langle h(\mathbf{k},t)h(\mathbf{k}',t')\rangle$, that describes the
shape fluctuations of a membrane is easily calculated~\cite{Seifert:1997} 
\begin{equation}
  \begin{split}
    \langle
    h(\mathbf{k},t)&h(\mathbf{k}',t')\rangle=\\&\exp\left[-\frac{\kappa
        k^3+\sigma k}{4\eta}|t-t'|\right]\frac{k_BT\,A}{\kappa k^4+\sigma k^2}\delta_{\mathbf{k},-\mathbf{k}'}.
  \end{split}
  \label{eq:correlation}
\end{equation}
\section{Diffusion on curved surfaces}
\label{sec:diff_curve}
When describing the dynamics of a particle free to diffuse laterally along the
membrane one has to bear in mind that the membrane is not flat but curved. The
dynamics of the particle is adequately expressed either by a Smoluchowski
equation that describes the time evolution of the probability of finding the
particle at a certain position, or by a Langevin equation that represents the
equation of motion of the particle position. In the following two sections
both dynamic equations appropriate for a protein diffusing freely on a curved
membrane the shape of which is given in the Monge gauge, will be introduced.
\subsection{Smoluchowski equation}
\label{sec:smolouchovsky}
In Cartesian coordinates the Smoluchowski equation for a particle diffusing
freely and isotropically on a flat $(x,y)$-plane is given by
\begin{equation}
  \frac{\partial P'(x,y,t)}{\partial t}=D\Delta P'(x,y,t),
  \label{eq:Smolouchovski_flat}
\end{equation}
with the diffusion coefficient $D$ and the probability $P'(x,y,t)dx\,dy$ of
finding the particle in the area element $dx\,dy$ at position $(x,y)$.  The
probability is normalized such that $\int_A\!dx\,dy\,P'(x,y,t)=1$ applies.  On
a curved surface the Laplace operator $\Delta$ needs to be replaced by the
Laplace-Beltrami operator, which is a function of the metric of the surface $g$
and the inverse metric tensor $g^{ij}$~\cite{Aizenbud:1982,Raible:2004}. The
resulting Smoluchowski equation takes the form
\begin{equation}
  \frac{\partial \mathcal{P}(x,y,t)}{\partial
    t}=D\sum_{i,j}\frac{1}{\sqrt{g}}\partial_i\sqrt{g}g^{ij}\partial_j\mathcal{P}(x,y,t).
  \label{eq:Smolouchovki_wrongnorm}
\end{equation}
The summations are to be taken over $x$ and $y$.
In the Monge gauge the metric of the surface is given by
\begin{equation}
  g\equiv1+h_x^2+h_y^2,
\end{equation}
with $h_x\equiv\partial_xh(x,y)$, for other subscripts accordingly, and the
inverse metric tensor by
\begin{equation}
  g^{ij}\equiv\frac{1}{g}\left(\begin{array}{cc}
      1+h_y^2&-h_xh_y\\
      -h_xh_y& 1+h_x^2
    \end{array}
  \right).
\end{equation}
The probability $\mathcal{P}(x,y,t)$ in eq.~\eqref{eq:Smolouchovki_wrongnorm}
is normalized such that $\int_A\!dx\,dy\,\sqrt{g}\,\mathcal{P}(x,y,t)=1$ is
valid. Assuming that in experiments typically the path of a particle projected
on the flat $(x,y)$-plane is regarded, it makes more sense to evaluate the
probability $P(x,y,t)\equiv\frac{1}{\sqrt{g}}\mathcal{P}(x,y,t)$ for which
the normalization $\int_A\!dx\,dy\,P(x,y,t)=1$ applies. To compare results
that neglect the curvature of the membrane with those that take it into
account correctly, the differences between $P'(x,y,t)$ and $P(x,y,t)$ need to
be analyzed. The Smoluchowski equation for $P(x,y,t)$ is
\begin{equation}
  \frac{\partial P(x,y,t)}{\partial
    t}=D\sum_{i,j}\partial_i\sqrt{g}g^{ij}\partial_j\frac{1}{\sqrt{g}}P(x,y,t).
  \label{eq:Smolouchovky_propernorm}
\end{equation}
The probability of finding the projected position of a particle within the
area element $dx\,dy$ around position $(x,y)$ is now given by $P(x,y,t)dx\,dy$.
\subsection{Langevin equation}
\label{sec:langevin}
To simulate the movement of a particle in a membrane it is more convenient to
use a Langevin equation which describes diffusion on a curved surface. In
general it is possible that several different Langevin equations produce the
same dynamics of the probability distribution $P(x,y,t)$. In the following we
will develop a realization of a Langevin equations in the Ito calculus that
leads to the dynamics of eq.~\eqref{eq:Smolouchovky_propernorm}. Within the
Monge description the Langevin equation we wish to develop is ideally the
equation of motion of the projected position $\mathbf{R}$ of the particle. The
actual particle position is then given by the vector $(X,Y,h(X,Y,t))$.  The
general form of the Langevin equation in Cartesian coordinates
is~\cite{Risken:1996}
\begin{equation}
  \frac{\partial R_i}{\partial t}=Db_i+G_{ij}\Gamma_j,
  \label{eq:Langevin}
\end{equation}
where $Db_i$ is the drift term that resembles an external force acting on the
particle, and $G_{ij}\Gamma_j$ is a stochastic force with
\begin{eqnarray}
  \langle\Gamma_i(t)\rangle&=&0\\
  \langle\Gamma_i(t)\Gamma_j(t')\rangle&=&2\delta_{ij}\delta(t-t').
\end{eqnarray}
For the curved system we will develop Langevin equations that obey the form of
eq.~\eqref{eq:Langevin} but where the information on the shape of the surface
is incorporated into the drift term $b_i$ and the strength $G_{ij}$ of the
stochastic force.  

The most general form of a Fokker-Planck equation in two-dimensional Cartesian
space is given by~\cite{Risken:1996}
\begin{equation}
  \frac{\partial P(x,y,t)}{\partial
    t}=\left[-\partial_iD^{(1)}_i+\partial_i\partial_jD^{(2)}_{ij}\right]P(x,y,t),
  \label{eq:FP}
\end{equation}
with the drift vector $D^{(1)}_i$ and the diffusive tensor
$D^{(2)}_{ij}$. Comparing this equation with
eq.~\eqref{eq:Smolouchovky_propernorm} we can identify 
\begin{eqnarray}
  \label{eq:line1}
  D^{(1)}_i&=&D\frac{1}{\sqrt{g}}\left(\partial_j(\sqrt{g}g^{ij})\right)\\
  D^{(2)}_{ij}&=&Dg^{ij}.
  \label{eq:line2}
\end{eqnarray}
Note that the partial derivative in eq.~\eqref{eq:line1} is not applied to
$P(x,y,t)$. If we derive the Langevin equations within the Ito calculus the
following relationships between the parameters $D^{(1)}_i$ and $D^{(2)}_{ij}$
of the Fokker-Planck equation~\eqref{eq:FP} and the drift term $Db_i$ and the
strength $G_{ij}$ of the stochastic force of eq.~\eqref{eq:Langevin} need to
be fulfilled~\cite{Risken:1996,Raible:2004}:
\begin{eqnarray}
  G_{ij}&=&\left(D^{(2)^{1/2}}\right)_{ij}=\left(D^{(2)^{1/2}}\right)_{ji}\\
  Db_i&=&D^{(1)}_i.
\end{eqnarray}
Using these relations and the identifications from eqs.~\eqref{eq:line1} and
\eqref{eq:line2} we arrive at the following Langevin equation
\begin{eqnarray}
  \label{eq:Ito_X}
  \frac{\partial X(t)}{\partial
    t}&&=\nonumber\\D\frac{1}{g^2}\!\!\!&&\!\!\!\left[2h_x^2h_yh_{xy}-h_xh_{xx}\left(1+h_y^2\right)-h_xh_{yy}\left(1+h_x^2\right)\right]\nonumber\\
  &&+\sqrt{D}\frac{1}{g-1}\left(\frac{h_x^2}{\sqrt{g}}+h_y^2\right)\Gamma_x\\
  &&+\sqrt{D}\frac{1}{g-1}h_xh_y\left(\frac{1}{\sqrt{g}}-1\right)\Gamma_y,\nonumber\\
  \label{eq:Ito_Y}
  \frac{\partial Y(t)}{\partial
    t}&&=\nonumber\\D\frac{1}{g^2}\!\!\!&&\!\!\!\left[2h_xh_y^2h_{xy}-h_yh_{xx}\left(1+h_y^2\right)-h_yh_{yy}\left(1+h_x^2\right)\right]\nonumber\\
  &&+\sqrt{D}\frac{1}{g-1}h_xh_y\left(\frac{1}{\sqrt{g}}-1\right)\Gamma_x\\
  &&+\sqrt{D}\frac{1}{g-1}\left(\frac{h_y^2}{\sqrt{g}}+h_x^2\right)\Gamma_y.\nonumber
\end{eqnarray}
Surprisingly, these equations comprise a drift term that is
induced by the curvature of the membrane and does not appear in the
Langevin equations for free diffusion on a flat plane. 
\section{Free diffusion within the preaveraging approximation}
\label{sec:analytical}
Before we turn to the simulation scheme we will give a short introduction to
our previous analytical calculations~\cite{Reister:2005} with which we
determine the measured, or projected, diffusion coefficient of a protein
diffusing freely in a membrane.

The solution of the Smoluchowski equation~\eqref{eq:Smolouchovky_propernorm}
that describes the time evolution of the probability distribution of the
projected position of the diffusing particle, is non trivial because the
prefactors containing the metric and the inverse metric tensor are time
dependent. If the time $\tau_\upxi$ it takes a particle to diffuse the length
$\upxi$ is much longer than the characteristic time $\tau_{\text{memb},\upxi}$
of membrane fluctuations with wavelength $\upxi$ we may apply a preaveraging
approximation, i.e., we may replace the time dependent prefactors in
eq.~\eqref{eq:Smolouchovky_propernorm} that contain partial derivatives of
$h(\mathbf{r},t)$ by their thermal averages. The membrane time scale is given
by the correlation time in eq.~\eqref{eq:correlation} as
$\tau_{\text{memb},\upxi}=\eta\upxi^3/(2\pi^3\kappa)$, while the diffusive
timescale is simply given as $\tau_\upxi=\upxi^2/4D$. Comparing these
timescales reveals that as long as lengths with
\begin{equation}
  \upxi \ll \pi^3\kappa/(2D\eta)
  \label{eq:preav}
\end{equation}
are regarded the preaveraging approximation should be valid.

When we average over the prefactors of eq.~\eqref{eq:Smolouchovky_propernorm}
we find that most terms vanish and the Smoluchowski equation simplifies
considerably. Taking into account only leading order prefactors it reads:
\begin{multline}
  \frac{\partial P(x,y,t)}{\partial
    t}=D\left[\left\langle\frac{1+h_y^2}{g}\right\rangle\frac{\partial^2
      P(x,y,t)}{\partial x^2}\right.\\+\left.\left\langle\frac{1+h_x^2}{g}\right\rangle\frac{\partial^2
      P(x,y,t)}{\partial y^2}\right].
\end{multline}
In an isotropic membrane the two remaining thermal averages are both equal and
therefore the Smoluchowski equation takes the form applicable for diffusion
on a flat surface, cf.~\eqref{eq:Smolouchovski_flat}, but now with a new
diffusion coefficient $D_{\text{proj}}$. The ratio of  $D_{\text{proj}}$
that would be measured in experiments, and the actual intramembrane
coefficient $D$ is given by:
\begin{equation}
  \frac{D_{\text{proj}}}{D}=\frac{1}{2}\left(1+\left\langle\frac{1}{g}\right\rangle\right).
  \label{eq:Dproj_D_analytical}
\end{equation}
By use of the relation $1/g=\int_0^{\infty}d\alpha\,\exp[-\alpha g]$ and the
Helfrich Hamiltonian~\eqref{eq:Helfrich} this ratio becomes:
\begin{multline}
  \frac{D_{\text{proj}}}{D}=\\\frac{1}{2}+\frac{1}{2}\int_0^{\infty}\!\!\!d\alpha\,\exp\Biggl[-\frac{1}{2}
  \sum_{\genfrac{}{}{0pt}{}{k_x,k_y} 
    {|\mathbf{k}|<q_m}}
  \!\!\ln\left(1+\frac{2\alpha}{\beta L^2}
    \frac{1}{\kappa k^2+\sigma}\right)\Biggr].
  \label{eq:Dproj_D_helfrich}
\end{multline}
A cutoff wavenumber $q_m\sim a$ needs to be introduced that is proportional to
the inverse of the microscopic length scale $a$ in the system. As we will see
later, in the simulations this microscopic length scale corresponds to the
lattice spacing. For a more detailed discussion of the numerical evaluation
and the results of this equation we refer the reader to our previous
work~\cite{Reister:2005}.

Instead of regarding the preaveraging approximation in the Smoluchowski
equation it may also be applied in the Langevin
equations~\eqref{eq:Ito_X},~\eqref{eq:Ito_Y}. The averaging process leads to a
vanishing drift term; the calculation of the the mean square displacement and
the subsequent derivation of the effective diffusion coefficient leads to
$D_{\text{proj}}/D=\langle G^2_{xx}+G^2_{yy}\rangle/2D=(1+\langle1/g\rangle)/2$. This is the
same result as eq.~\eqref{eq:Dproj_D_analytical}.
\section{Simulation scheme}
\label{sec:simulation}

The analytical calculations in the last section were performed within a
preaveraging approximation. In the present work we do not apply this
approximation, but rather simulate the equation of motion for the
membrane~\eqref{eq:eom_r} and the Langevin equations~\eqref{eq:Ito_X}
and~\eqref{eq:Ito_Y} for the protein movement.  The simulation scheme that we
will introduce during the next paragraphs resembles the effective evaluation
of this coupled set of dynamic equations by means of discrete numerical
integration in time.

First let us turn to the numerical evaluation of the time evolution of the
membrane shape. If we regard eq.~\eqref{eq:eom_k} and assume periodic boundary
conditions it is obvious that the numerical integration is most effectively
implemented in Fourier space because the evolution of the height function modes
$h(\mathbf{k},t)$ takes place independently of each other. If the diffusion
of the protein were not taken into account the whole time evolution of the
membrane shape could be performed in $\mathbf{k}$-space. However, the real
space representation $h(\mathbf{r},t)$ becomes necessary to develop the
position of the protein. Due to the periodic boundary conditions
$\mathbf{k}$-vectors are of the form $\mathbf{k}=2\pi(l,m)/L$ with $A=L^2$.
Because in real space $h(\mathbf{r},t)$ is expressed on a quadratic lattice of
$N\times N$ lattice sites the restriction $-N/2<l,m\leqslant N/2$ applies. The
lattice spacing $a$ is given by $a\equiv L/N$.

When numerically implementing eq.~\eqref{eq:eom_k} one must remember that
$h(\mathbf{k},t)$ may be a complex number with a real part $h_r(\mathbf{k},t)$
and an imaginary part $h_i(\mathbf{k},t)$ such that
$h(\mathbf{k},t)=h_r(\mathbf{k},t)+ih_i(\mathbf{k},t)$. Both for the real and
the imaginary part an equation of motion is necessary.  The numerical
equations that are used in the simulations to develop the membrane shape during
a timestep $\Delta t$ are of the form
\begin{multline}
  h_{r/i}(\mathbf{k},t)=h_{r/i}(\mathbf{k},t-\Delta t)\\+\Delta t \frac{1}{4\eta
    k}\left[\kappa k^4+\sigma k^2\right]h_{r/i}(\mathbf{k},t-\Delta t)\\+\sqrt{\lambda k_BT
    A\,\Delta t\,/(4\eta k)}\,r.
  \label{eq:eqm_h_k}
\end{multline}
The random number $r$ is Gaussian and therefore $\langle r \rangle=0$ and
$\langle r^2 \rangle=1$ applies.  The factor $\lambda$ in the random term is
either $1$ or $2$ as we will now explain: Due to $h(\mathbf{r},t)$ and
$\xi(\mathbf{r},t)$ being real quantities not all modes have an imaginary
part. In particular modes with wave vectors $\mathbf{k}=2\pi(l,m)/L$ with
$(l,m)=(0,0),\,(0,N/2),\,(N/2,0),\,(N/2,N/2)$ are purely real, while all
others are complex. In order to fulfill the fluctuation dissipation theorem
from eq.~\eqref{eq:fluct_diss_k} the four purely real modes only have an
equation of motion for the real part with $\lambda=2$, while all other
independent modes have two equations of motion, one for the real and one for
the imaginary part, each with $\lambda=1$. The real and the imaginary part of
$\xi(\mathbf{k,t})$ are assumed not to be correlated. Note that not all modes
are independent, because $h(\mathbf{r},t)$ is a real function. Only a set of
independent modes is updated via~\eqref{eq:eqm_h_k}, while the dependent modes
are set such that $h(\mathbf{k},t)=h^*(-\mathbf{k},t)$.

Regarding the random term in eq.~\eqref{eq:eqm_h_k} it becomes evident that it
diverges for $k=0$ due to the Onsager coefficient $\Lambda(\mathbf{k})$. The
mode $h(\mathbf{k}=0,t)$ is a measure for the distance between the center of
mass of the membrane and the flat $(x,y)$-plane.  Therefore, fluctuations of
$h(\mathbf{k}=0,t)$ just describe a movement of the membrane as a whole. Such
movement of the center of mass is of no relevance for the membrane and
diffusive properties of interest in this work, so we keep
$h(\mathbf{k}=0,t)=0$ fixed at all times.

So far the simulation scheme is rather similar to the \emph{Fourier space
  Brownian dynamics} method introduced earlier by Lin and
Brown~\cite{Lin:2004,Lin:2004a,Lin:2005}. But in our simulations we
additionally take into account the diffusion of a freely diffusing particle
along the curved surface given by the membrane shape. After an update in the
membrane shape using eqs.~\eqref{eq:eqm_h_k} we will now update the position
of the diffusing particle by using a discrete version of eqs.~\eqref{eq:Ito_X}
and \eqref{eq:Ito_Y}~\cite{Risken:1996,Raible:2004}:
\begin{multline}
  X(t+\Delta t)=
  X(t)\\+D\frac{1}{g^2}\left[2h_x^2h_yh_{xy}-h_xh_{xx}\left(1\!+\!h_y^2\right)-h_xh_{yy}\left(1\!+\!h_x^2\right)\right]\Delta
  t\\
  +\sqrt{D}\frac{1}{g-1}\left(\frac{h_x^2}{\sqrt{g}}+h_y^2\right)\sqrt{2\Delta
    t}\,r_x\\
  +\sqrt{D}\frac{1}{g-1}h_xh_y\left(\frac{1}{\sqrt{g}}-1\right)\sqrt{2\Delta
    t}\,r_y.
  \label{eq:Ito_X_discrete}
\end{multline}
For $Y(t)$ the corresponding equation is valid. Hereby we use
$h_x(\mathbf{R}(t),t)$, for all other partial derivatives of $h$ accordingly,
at the position of the particle $\mathbf{R}(t)$ at time $t$. The random
numbers $r_x$ and $r_y$ are again Gaussian with $\langle r_i\rangle=0$ and
$\langle r_ir_j\rangle=\delta_{ij}$.  This equation clearly describes an
off-lattice movement of the particle. It is also conceivable to simulate the
protein movement through a random walk on the lattice used for the membrane.
In such a scheme we would check the probability of a particle moving to a
neighboring lattice site within a timestep $\Delta t$ and then decide whether
the particle is to hop.  In biological systems, however, the time it takes a
particle to diffuse the length of a lattice spacing $a$ in the simulation is
much larger than the typical timescale of membrane fluctuations with wave
length $a$. This would typically lead to significant changes in the membrane
shape before a particle jump is successful. Therefore, the off-lattice version
of the particle's random walk is much more favorable, although the partial
derivatives of $h$ need to be extrapolated to the position of the particle.
This is realized by the following procedure: Multiplying $h(\mathbf{k},t)$
with the appropriate $\mathbf{k}$-vectors and subsequently performing the
Fourier backtransform, we arrive at the first and second partial derivatives
of $h(\mathbf{r},t)$ with respect to $x$ and $y$.  All discrete Fourier
transformations necessary for the simulations are effectively implemented
using the FFTW routines~\cite{FFTW05}.  Assume that the protein is to be found
somewhere between the four lattice sites $(i,j)$, $(i+1,j)$, $(i,j+1)$, and
$(i+1,j+1)$, with $0\leqslant i,j\leqslant N-1$. The quantity $A(\mathbf{R})$
to be extrapolated to the particle position is given at the lattice sites by
$A(i,j)$, $A(i+1,j)$, etc. The distance between the particle and the line
connecting $(i,j)$ and $(i,j+1)$ is to be $\mu a$ and the distance between
particle and the line connecting $(i,j)$ and $(i+1,j)$ is $\nu a$. The
linearly extrapolated value of $A$ at the particle position
$\mathbf{R}=(a(i+\mu),a(j+\nu))$ is calculated by use of:
\begin{multline}
  A(\mathbf{R})=A(i,j)\\-\nu a(A(i,j)-A(i,j+1))-\mu a(A(i,j)-A(i+1,j))\\+\mu\nu a^2(A(i,j)-A(i+1,j)\\+A(i+1,j+1)-A(i,j+1))
\end{multline}
After all the necessary quantities have been determined at position
$\mathbf{R}$ the new particle position is calculated with
eq.~\eqref{eq:Ito_X_discrete}. During the timestep of course also the shape of
the membrane will change. This is accounted for by employing
eq.~\eqref{eq:eqm_h_k} in the next computational step to get
$h(\mathbf{k},t+\Delta t)$. Then we again update the particle position and so
on. Repeating the two step process of membrane shape update and particle
movement makes out our simulation scheme.

Instead of simulating the diffusion of one particle in one membrane it
possible to insert several particles into the membrane that do not interact
with each other. For each particle a separate Langevin equation needs to
evaluated, but only one membrane equation of motion. Since the Fourier
transform of the membrane configuration is the most time consuming element of
the code, the insertion of more than one particle saves computing
time. However, the average distance between the particles should be
sufficiently large in order for the particle paths to be independent.  
Results presented in sections~\ref{sec:sim_diff} and~\ref{sec:sim_corr}
follow from averaging over 2500 paths in 100 different membranes. In each
independent membrane 25 particles, that do not interact with each other, are
allowed to diffuse.

All simulation results presented in this paper were performed on a
$50\times50$ lattice. If we assume (arbitrarily) that the
lattice spacing $a$ corresponds to 10nm the system size is $L=0.5\mu$m. 
The viscosity $\eta$ of the
water surrounding the membrane is $\eta=10^{-9}$Js/cm$^3$. With the above
chosen lattice spacing $a$ and the temperature $T=300$K, the viscosity used
in the Onsager coefficient of the simulations is
$\eta=2.4\times10^{-7}k_BT\text{s}/a^3$.

The choice of the appropriate timestep $\Delta t$ is determined by two
timescales, namely the smallest timescale of membrane fluctuations
$\tau_{\text{memb,min}}$ and the time $\tau_{\text{diff},a}$ it takes a
particle on average to diffuse a lattice spacing $a$. Only if $\Delta t$ is
significantly smaller than $\tau_{\text{memb,min}}$ then the evolution of the
membrane shape will be numerically stable.  The membrane timescale is given by
$\tau_{\text{memb,min}}=4\eta/(\kappa k_{\text{max}}^3+\sigma
k_{\text{max}})$, cf.~eq.~\eqref{eq:correlation}. The maximum wave number
$k_{\text{max}}=\sqrt{2}\pi/a$ is determined by the minimal microscopic length
scale of the system, i.e., in simulations the lattice spacing $a$.
Additionally the average length a particle moves during $\Delta t$ needs to be
much shorter than $a$, because we regard the membrane shape only right at the
beginning of the jump. If the jump is too big, the actual particle path along
the membrane is not taken into account with the necessary accuracy. The
diffusive timescale is determined by $\tau_{\text{diff},a}=a^2/4D$.  To
perform the simulations $\Delta t$ should be considerably smaller than both
$\tau_{\text{memb,min}}$ and $\tau_{\text{diff},a}$.  Timesteps $\Delta t$
used in the presented simulations range from $5\times10^{-10}$s to
$2\times10^{-9}$s. Typical simulation runs comprise $\sim2\times10^6$
timesteps which took approximately 30-35 minutes on a 64 node Beowulf cluster
with Pentium IV, 3.2 GHz processors.
\section{Results}
\label{sec:results}
\begin{figure}
  \begin{center}
    \includegraphics[width=0.9\linewidth,clip]{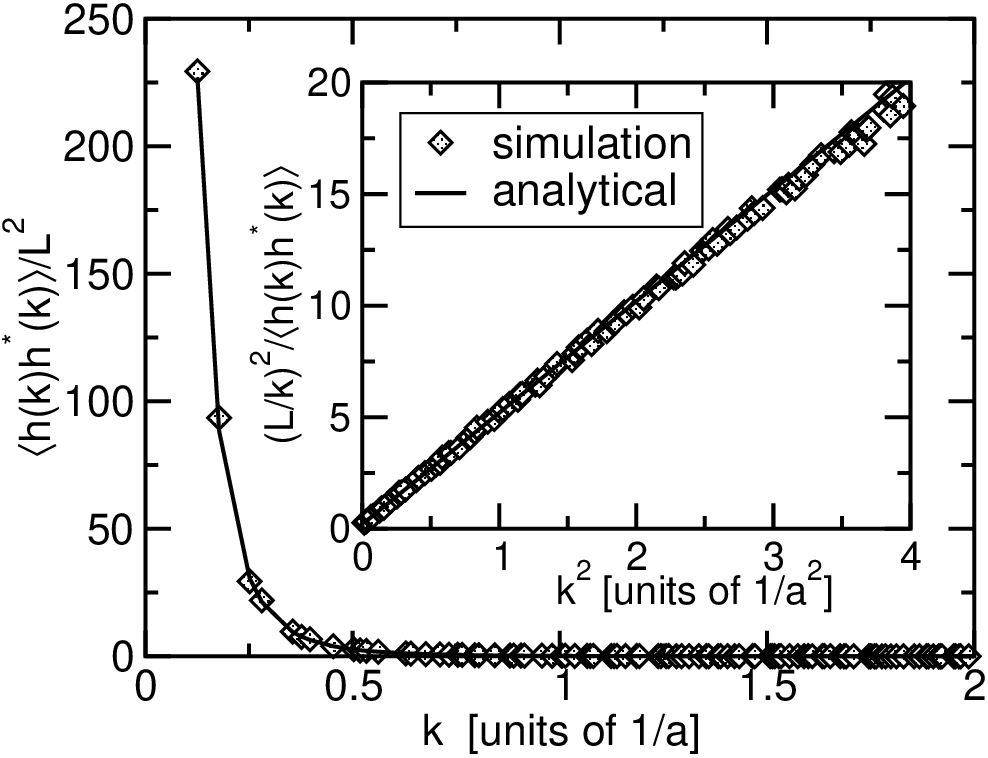}
  \end{center}
  \caption{\label{fig:hqhq_kappa5_sigmaLL500}Height-height correlation function
    $\langle h(\mathbf{k})h^*(\mathbf{k})\rangle$ as a function of wave number
    $k$ (in units of $1/a$). To illustrate the agreement between the
    analytical and the simulation result for larger $k$ values, we plot
    $k^2/\langle h(\mathbf{k})h^*(\mathbf{k})\rangle$ as a function of $k^2$
    in the inset.  The simulation results, symbolized by the diamond symbols,
    were obtained by averaging over 250 independent membrane configurations
    created during runs with dimensionless physical parameters $\beta\kappa=5$
    and $\beta\sigma L^2=500$ on a 50$\times$50 lattice. The numerical
    timestep was $\Delta t=1.7\times10^{-10}$s.}
\end{figure}
\subsection{Validation of membrane fluctuations}
After developing a simulation code it is always necessary to test it by
comparing its results with results previously obtained with another method. In
this section we will show that the evolution of the membrane shape results in
the membrane fluctuations given by eq.~\eqref{eq:correlation} that follow
analytically for a membrane with the Helfrich energy~\eqref{eq:Helfrich} and
the Onsager coefficient of eq.~\eqref{eq:Onsager}.

In fig.~\ref{fig:hqhq_kappa5_sigmaLL500} we display the equal time correlation
function $\langle h_r^2(\mathbf{k},t)+h_i^2(\mathbf{k},t)\rangle$ as a
function of $k\equiv|\mathbf{k}|$ for  
$\beta\kappa=5$ and $\beta\sigma L^2=500$. This corresponds to a surface
tension on the order of $8\times10^{-3}$mJ$/$m$^2$. 
\begin{figure}
  \begin{center}
    \includegraphics[width=0.9\linewidth,clip]{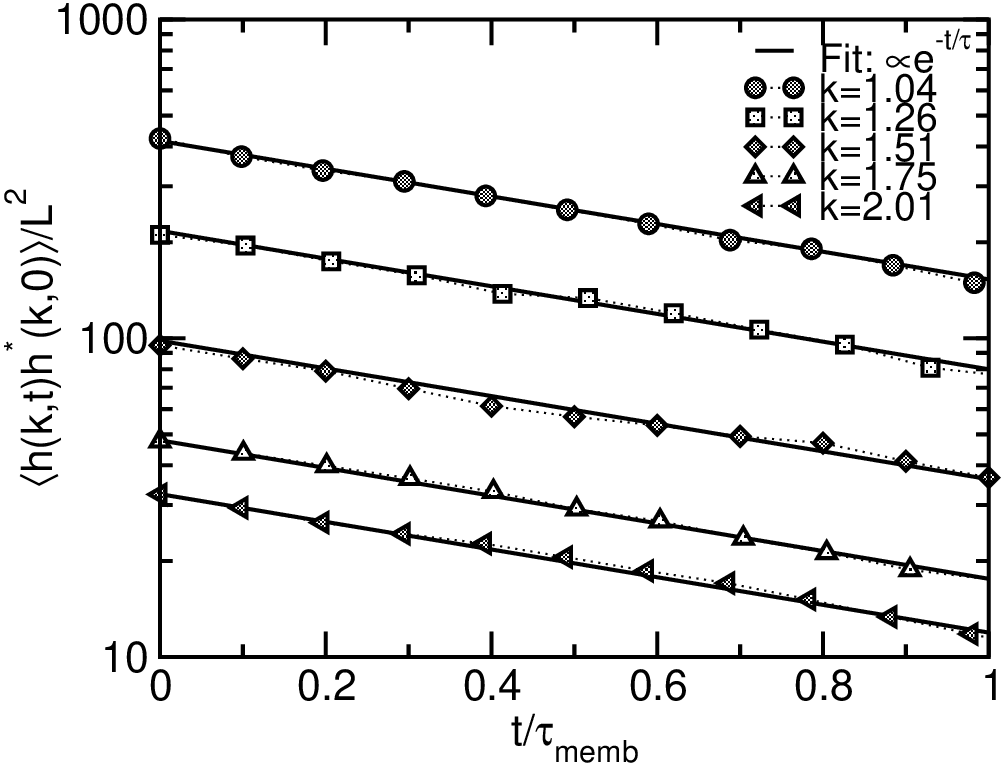}
  \end{center}
  \caption{\label{fig:hq0hqt_kappa5_sigmaLL500}Time correlation function 
    $\langle h(\mathbf{k},t)h^*(\mathbf{k},0)\rangle$ as a function of the
    dimensionless time $t/\tau_{\text{memb}}(k)$ for the given values of $k$.
    The correlation time $\tau_{\text{memb}}(k)=4\eta/(\kappa k^3+\sigma k)$
    is given by eq.~\eqref{eq:correlation}. Symbols are results from the
    simulations, while straight lines resemble fits $\propto
    \exp[-t/\tau_{\text{memb}}]$.}
\end{figure}
Typical values for the dimensionless bending rigidity $\beta\kappa$ of lipid
bilayer membranes are between 1 and 50.

If we compare the simulation results with the analytical result that is given
by the solid line we see a good agreement for small wave numbers. For higher
$k$ values it is more convenient to plot $k^2/\langle
h_r^2(\mathbf{k},t)+h_i^2(\mathbf{k},t)\rangle$ as a function of $k^2$, which
is done in the inset. The linear behavior expected from the analytical
calculations is well reproduced by the simulations.

In order to test the time correlations of the membrane fluctuations we plot
the correlation function $\langle h(\mathbf{k},t)h^*(\mathbf{k},0)\rangle$ as
a function of the dimensionless time $t/\tau_{\text{memb}}(k)$ for several
randomly chosen wave numbers in fig.~\ref{fig:hq0hqt_kappa5_sigmaLL500}.  The
same parameters as in fig.~\ref{fig:hqhq_kappa5_sigmaLL500} were used.  The
exponential fits $\propto\exp[-t/\tau_{\text{memb}}]$ to the simulation
results reveal a good agreement between the analytical correlation times and
the results obtained from our numerical membrane update scheme.
\subsection{Free membrane-bound diffusion}
\label{sec:sim_diff}
In the previous section we showed that our simulation scheme reproduces
membrane fluctuations correctly.  In this section we show that it is also
capable of adequately describing the diffusion of a protein in the membrane.

To determine the projected diffusion coefficient in the simulations we use the
relation $\langle(\mathbf{R}(t)-\mathbf{R}(0))^2\rangle=4D_{\text{proj}}t$.
The mean square displacement $\langle\Delta
\mathbf{R}^2(t)\rangle\equiv\langle(\mathbf{R}(t)-\mathbf{R}(0))^2\rangle$ as
a function of time is determined from the simulations by averaging over a
large number of independent particle paths.  
\begin{figure}
  \begin{center}
    \includegraphics[width=0.9\linewidth,clip]{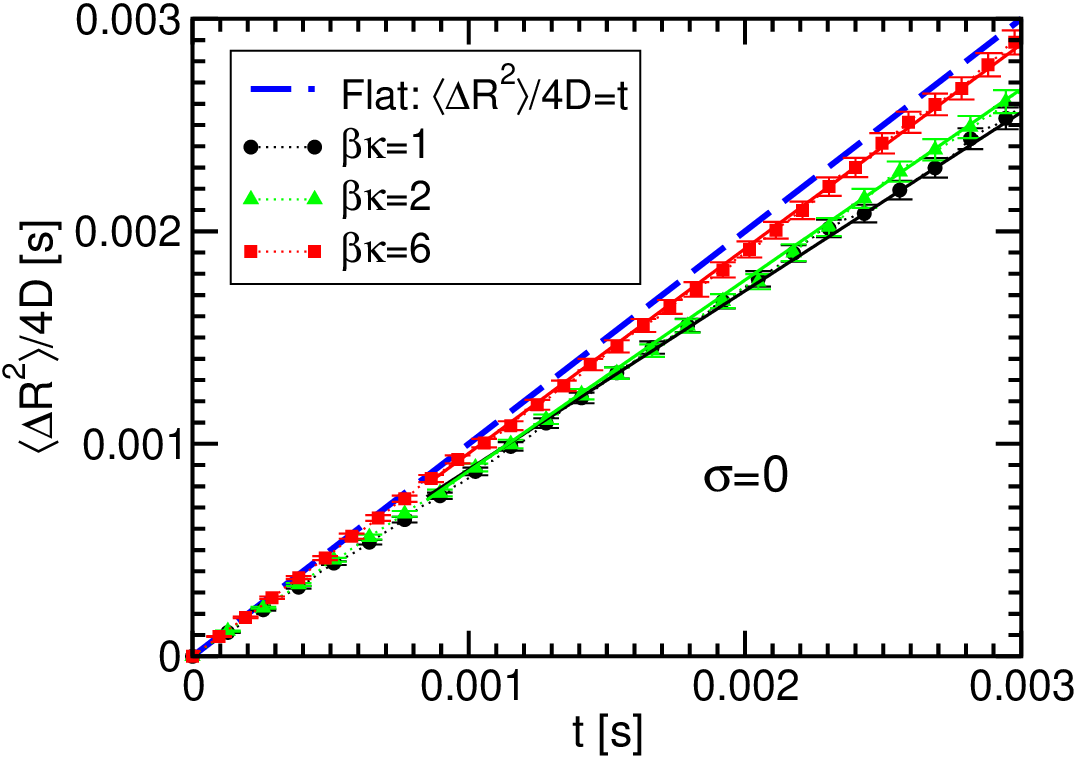}
  \end{center}
  \caption{\label{fig:linear}(Color online) Elucidation how $D_{\text{proj}}$ is determined
    from simulation results (symbols): the averaged mean square displacement
    $\langle\Delta R^2(t)\rangle/4D$ is plotted as a function of time. The
    error bars correspond to the standard error of averaging. The slopes of
    linear fits (solid lines) to these results determine the diffusion
    coefficient. The projected diffusion coefficient on a fluctuating membrane
    is always smaller than on a flat plane (thick dashed line). }
\end{figure}
In fig.~\ref{fig:linear} we show $\langle\Delta \mathbf{R}^2(t)\rangle/4D$ as
a function of time for the given values of $\beta\kappa$ and $\sigma=0$. The
chosen intramembrane diffusion coefficient is $D=10^5a^2/$s; this corresponds
to an experimental value of $D=10^{-7}$cm$^2/$s. The results evidently show a
linear increase of $\langle\Delta \mathbf{R}^2(t)\rangle$ with time.
Furthermore, the comparison of the simulation results with the line that gives
the expected mean square displacement if the membrane were flat displays that
the projected diffusion coefficient is smaller than the actual intramembrane
diffusion coefficient.  This is of course expected and also seen in the result
of eq.~\eqref{eq:Dproj_D_analytical}, because membrane fluctuations increase
the actual path of the protein.

\begin{figure}[t!]
  \begin{center}
    \includegraphics[width=0.9\linewidth,clip]{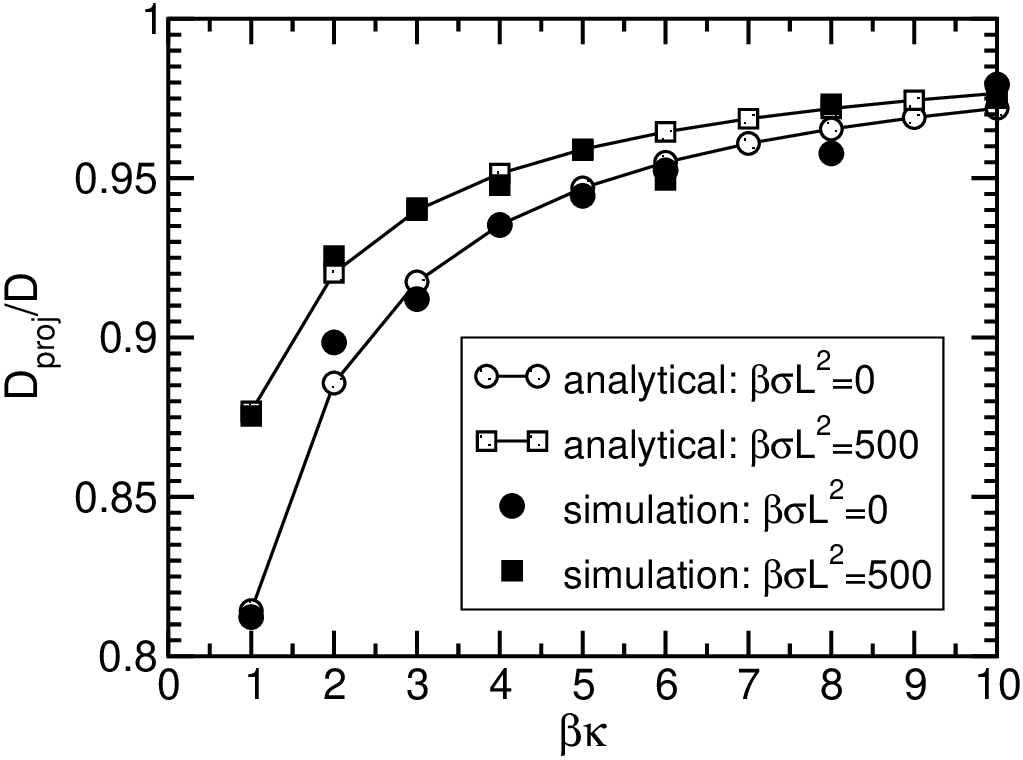}
  \end{center}
  \caption{\label{fig:result}Comparison of simulation and analytical results for
    $D_{\text{proj}}/D$ as a function of bending rigidity $\beta\kappa$ for
    the two given effective tensions. Good agreement is observed.}
\end{figure}

The slope of the linear fit to the simulation results some of which are
displayed in fig.~\ref{fig:linear} corresponds to the ratio
$D_{\text{proj}}/D$. In fig.~\ref{fig:result} we plot $D_{\text{proj}}/D$
resulting from the simulations and the numerical evaluation of the integral in
eq.~\eqref{eq:Dproj_D_helfrich} as a function of bending rigidity
$\beta\kappa$ both for vanishing tension and $\beta\sigma L^2=500$.  
Both simulations and preaveraging calculations show that for small bending
rigidity and small tension the difference between projected and actual
diffusion coefficient becomes more and more pronounced. This result is
plausible: decreasing tension $\sigma$ and bending rigidity $\kappa$ leads to
stronger membrane fluctuations. The difference between actual and projected
particle path becomes bigger and therefore the fluctuation effect is enhanced.
If the system size is increased, not shown here, this also leads to an
increase in fluctuation strength.  The comparison of simulation and analytical
results displays a very good overall agreement.

\begin{figure}
  \begin{center}
    \includegraphics[width=0.9\linewidth,clip]{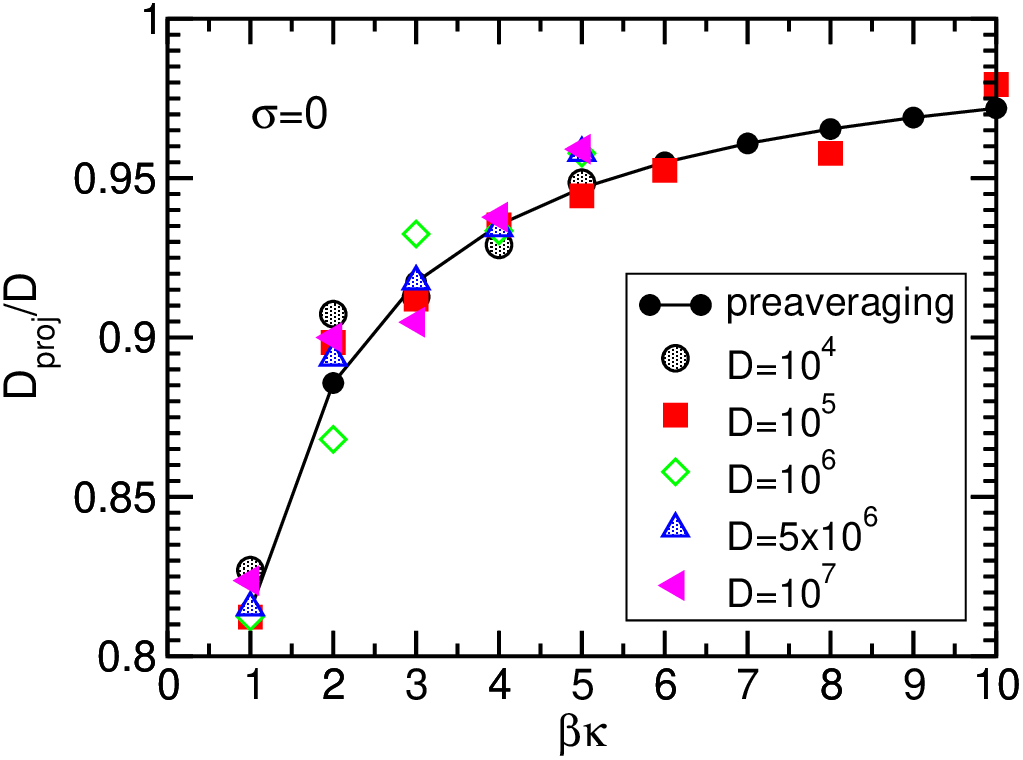}
  \end{center}
  \caption{\label{fig:Dproj_D_allD}(Color online) Ratio of the projected to the
    intramembrane diffusion coefficient $D_{\text{proj}}/D$ as a function of
    $\beta\kappa$ for various diffusion coefficients (given in units of
    $a^2$/s) and $\sigma=0$. Simulation results for all $D$ cannot be
    distinguished from the preaveraging result.}
\end{figure}

Results in fig.~\ref{fig:result} were achieved for a single diffusion
coefficient $D$. To study the influence of the intramembrane diffusion
coefficient we also perform simulations for five different coefficients
$D=10^4$, $10^5$, $10^6$, $5\times10^6$, $10^7a^2$/s and $\sigma=0$.  In
fig.~\ref{fig:Dproj_D_allD} we display the ratio $D_{\text{proj}}/D$ as a
function of $\beta\kappa$ for all used $D$ as derived from the average mean
square displacement $\langle\Delta\mathbf{R}^2(t)\rangle$ of the diffusing
particles. The results show that $D_{\text{proj}}/D$ is seemingly independent
of the diffusion coefficient $D$: for all values of $D$ a good agreement of
the simulation results is observed. Comparing the simulation results with the
analytical result invoking the preaveraging approximation we also find that
the agreement is very good.
But this is rather surprising: a priori the preaveraging approximation should
only be applicable if the time it takes a particle to diffuse the length
$\upxi$ is much larger than the correlation time of membrane fluctuations with
wavelength $\upxi$. This is expressed in eq.~\eqref{eq:preav} where we have a
crossover length scale
$\upxi_{\text{co}}\equiv\pi^3\kappa/(2D\eta)\simeq6\times10^7\beta\kappa/D$.
If diffusion on length scales below $\upxi_{\text{co}}$ is analyzed the
preaveraging approximation should lead to good agreement, while one would
expect to see a change for larger length scales. For the smallest regarded
diffusion coefficients $\upxi_{\text{co}}$ is on the order of $10^3a$. With a
system length of $L=50a$ all lengths in the system are below the crossover
length scale and therefore the agreement of the simulation results with the
preaveraging result, as seen in figs.~\ref{fig:result}
and~\ref{fig:Dproj_D_allD}, is expected. For the largest regarded $D$ the
crossover length is on the order of $10a$. Thus there are very many membrane
fluctuation modes in the system with wave lengths larger than
$\upxi_{\text{co}}$. In this case we would have expected the interplay of
membrane and diffusive timescales to be observable in the effective diffusion
coefficient.  However, fig.~\ref{fig:Dproj_D_allD} shows that this is not the
case. To understand why the preaveraging approximation leads to good agreement
with explicit simulations even for large $D$, it is necessary to study the
correlations of the drift term in the Langevin equations~\eqref{eq:Ito_X}
and~\eqref{eq:Ito_Y} that is caused by the metric of the system. This is the
subject of the following section where we discuss the validity of the
preaveraging approximation.
\subsection{Validity of the preaveraging approximation}
\label{sec:sim_corr}
The last section had the unexpected outcome that the analytical calculation
within the  preaveraging approximation describes particle diffusion well even
when the diffusion coefficient is so large that one would assume this
approximation to break down. In this section we study in more
detail the validity of the preaveraging approximation. 

\begin{figure}
  \begin{center}
    \includegraphics[width=0.9\linewidth,clip]{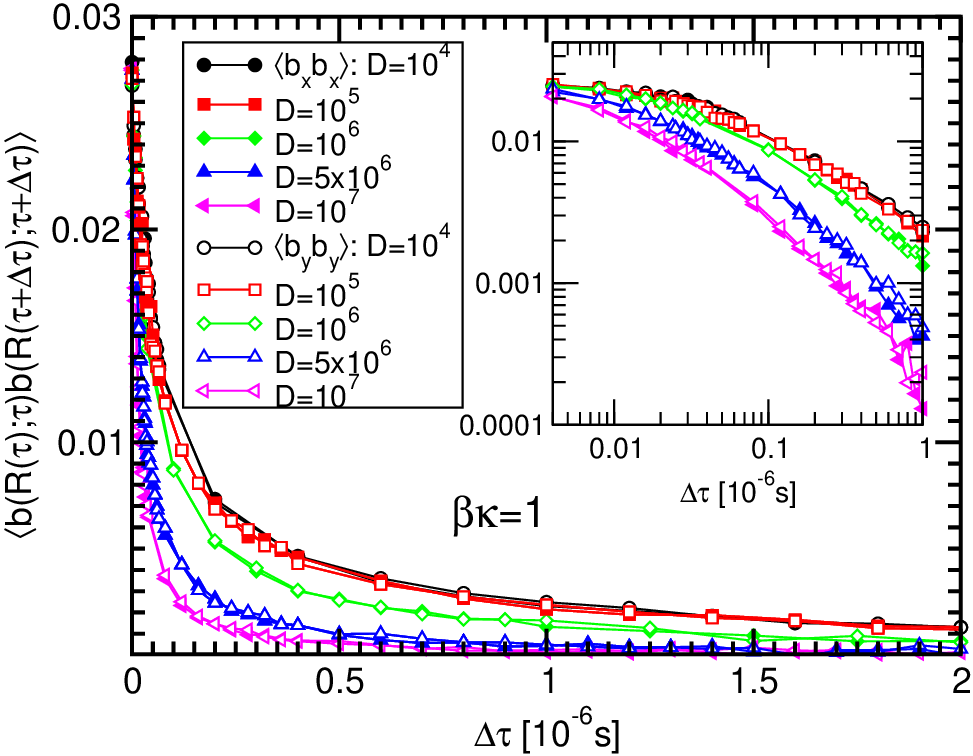}\\[1cm]
    \includegraphics[width=0.9\linewidth,clip]{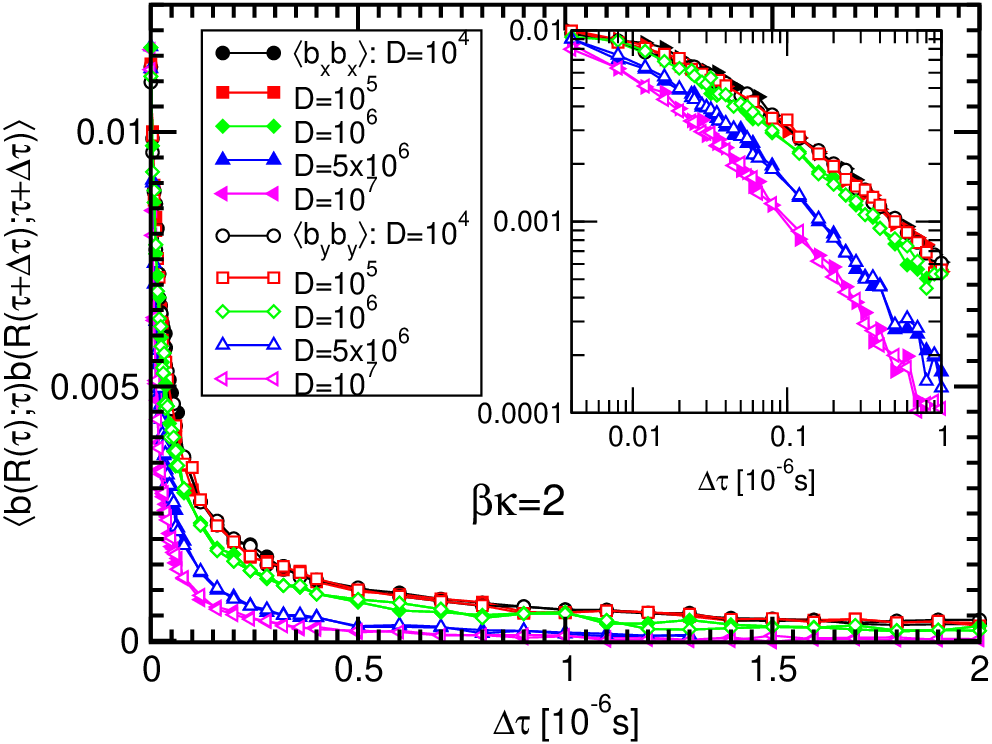}\\
  \end{center}
  \caption{\label{fig:correlation}(Color online) Correlation functions $\langle
    b_x(\mathbf{R}(\tau);\tau)b_x(\mathbf{R}(\tau+\Delta
    \tau);\tau+\Delta\tau)\rangle$ (solid symbols) and $\langle
    b_y(\mathbf{R}(\tau);\tau)b_y(\mathbf{R}(\tau+\Delta
    \tau);\tau+\Delta\tau)\rangle$ (white symbols) as a function of
    $\Delta\tau$ for $\beta\kappa=1$ (top panel) and $\beta\kappa=2$ (bottom
    panel). Different symbols apply for different diffusion coefficients
    (given in units of $a^2$/s). The insets display the same results as double
    logarithmic plots.}
\end{figure}

\begin{figure}
  \begin{center}
    \includegraphics[width=0.9\linewidth,clip]{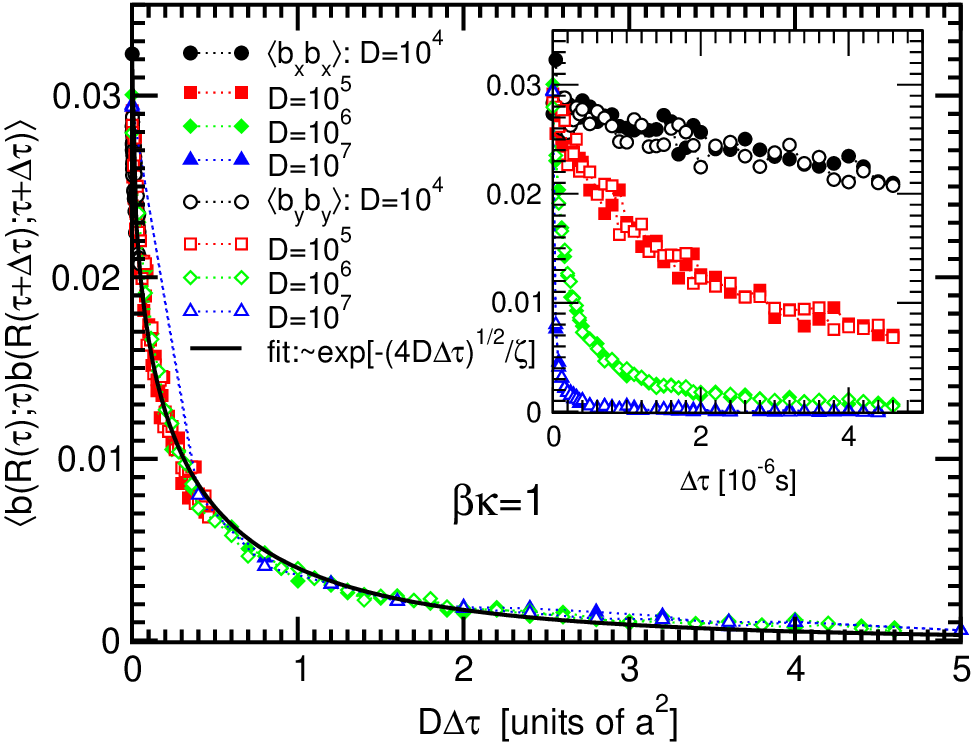}
  \end{center}
  \caption{\label{fig:correlation_stiff}(Color online) Correlation functions $\langle
    b_x(\mathbf{R}(\tau);\tau)b_x(\mathbf{R}(\tau+\Delta
    \tau);\tau+\Delta\tau)\rangle$ (solid symbols) and $\langle
    b_y(\mathbf{R}(\tau);\tau)b_y(\mathbf{R}(\tau+\Delta
    \tau);\tau+\Delta\tau)\rangle$ (white symbols) as a function of
    $D\Delta\tau$ and $\Delta\tau$ (inset) resulting from diffusion on 100
    different quenched membranes with 25 particles each for $\beta\kappa=1$.
    Different symbols apply for different diffusion coefficients.}
\end{figure}

To this end it is helpful to regard the Langevin equations~\eqref{eq:Ito_X}
and~\eqref{eq:Ito_Y} governing particle diffusion.  The diffusion coefficient
is defined through the mean square displacement of the diffusing particle.
Using eq.~\eqref{eq:Langevin} we can formally write the mean square
displacement as
\begin{multline}
  \langle\Delta\mathbf{R}^2(t)\rangle=\\D^2\int_0^{t}d\tau\int_0^td\tau'\left\{\sum_i\langle
    b_i(\mathbf{R}(\tau);\tau)b_i(\mathbf{R}(\tau');\tau')\rangle\right\}\\+2\sum_i\langle
  G_{ii}^2 \rangle t.
  \label{eq:msd_formal}
\end{multline}
The last term that is linear in time $t$ coincides with the preaveraging
result, because $\langle G^2_{xx}+G^2_{yy}\rangle/2D=(1+\langle1/g\rangle)/2$,
see eq.~\eqref{eq:Dproj_D_analytical}.  Hence the diffusion coefficient
derived through $\langle\Delta\mathbf{R}^2(t)\rangle/4t$ gets an additional
term caused by the correlations of the drift term. If we are interested in the
ratio of the projected and the actual diffusion coefficient it may be written
as the sum of two terms
$D_{\text{proj}}/D=D_{\text{proj,preav}}/D+D_{\text{proj,drift}}/D$, with the
additional contribution
\begin{multline}
  \frac{D_{\text{proj,drift}}}{D}=\\\frac{1}{2t}D\int_0^td\tau\int_0^{\tau}d\tau'\left\{\langle
    b_x[h(\mathbf{R}(\tau);\tau)]
    b_x[h(\mathbf{R}(\tau');\tau')]\rangle\right.+\\\left.\langle
    b_y[h(\mathbf{R}(\tau);\tau)]
    b_y[h(\mathbf{R}(\tau');\tau')]\rangle\right\},
  \label{eq:Dprojdrift_D}
\end{multline}
with 
\begin{equation}
  \begin{split}
    b_x[h]\!&\equiv\!\frac{1}{g^2}\left[2h_x^2h_yh_{xy}\!-\!h_xh_{xx}\left(1\!+\!h_y^2\right)\!-\!h_xh_{yy}\left(1\!+\!h_x^2\right)\right],\\
    b_y[h]\!&\equiv\!\frac{1}{g^2}\left[2h_y^2h_xh_{xy}\!-\!h_yh_{yy}\left(1\!+\!h_x^2\right)\!-\!h_xh_{xx}\left(1\!+\!h_y^2\right)\right].
  \end{split}
\end{equation}
In order to estimate this contribution we calculate the functions $\langle
b_x(\mathbf{R}(\tau);\tau)b_x(\mathbf{R}(\tau+\Delta
\tau);\tau+\Delta\tau)\rangle$ and $\langle
b_y(\mathbf{R}(\tau);\tau)b_y(\mathbf{R}(\tau+\Delta
\tau);\tau+\Delta\tau)\rangle$ from our simulation data for the five
previously regarded diffusion coefficients. Note that the correlations do not
only depend on the pure time interval $|\tau-\tau'|$ but also on the distance
$|\mathbf{R}(\tau)-\mathbf{R}(\tau')|$ the particle travels during this time
interval.  The correlation functions are displayed in the top and bottom
panels of fig.~\ref{fig:correlation} as a function of $\Delta\tau$ for
$\beta\kappa=1$ and $\beta\kappa=2$. In this whole section we set the
effective surface tension $\sigma=0$.
We find that the correlation function is overall smaller for larger
$\beta\kappa$ as is expected. This means that the larger the bending rigidity
$\kappa$ or the effective tension $\sigma$, not shown here, the less important
is the contribution caused by the drift term.  The expectation for an
isotropic system that the correlation functions $\langle
b_x(\mathbf{R}(\tau);\tau)b_x(\mathbf{R}(\tau+\Delta
\tau);\tau+\Delta\tau)\rangle$ and $\langle
b_y(\mathbf{R}(\tau);\tau)b_y(\mathbf{R}(\tau+\Delta
\tau);\tau+\Delta\tau)\rangle$ coincide, is also fulfilled. Furthermore, a
faster decrease of the correlation function with increasing diffusion
coefficient $D$ is observed.  The decrease of the correlation function with
time $\Delta\tau$ is determined by two processes: the change in membrane shape
and the movement of the particle along the membrane. If the particle did not
diffuse, i.e., $D=0$, the displayed correlation functions would still decrease
with increasing the time interval due to the evolution of the membrane shape.
This contribution to the decrease obviously does not depend on how fast the
particle diffuses within the membrane, but only depends on membrane parameters
like bending rigidity $\kappa$ or surface tension $\sigma$. In the other
extreme when membrane time scales $\tau_{\text{memb}}$ approach infinity, and
the particle diffuses on a fixed membrane shape, the decrease of the
correlation function is determined by the distance a particle travels during a
fixed time interval.  This is shown in fig.~\ref{fig:correlation_stiff} for
$\beta\kappa=1$, where we display the correlation functions $\langle
b_{i}(\mathbf{R}(\tau);\tau)b_{i}(\mathbf{R}(\tau+\Delta
\tau);\tau+\Delta\tau)\rangle$ for particles diffusing on different quenched
membranes over which we average afterwards.  The ``quenched'' configurations
used in the simulations are obtained by evolving the membrane shape
for such a long time that thermal equilibrium has been reached.
Regarding the correlation functions as a function of time we see that the
smaller the diffusion coefficient the slower the decrease of the correlations.
Multiplying $\Delta\tau$ with the diffusion coefficient leads
to a perfect match of all four lines.  This is a clear indication that the
correlation function depends only on the average distance
$\sqrt{4D\Delta\tau}$ a particle travels during a certain time. The thick
solid line is the result of an exponential fit
$\propto\exp(-\sqrt{4D\Delta\tau}/\zeta)$; the correlation length for
$\beta\kappa=1$ is given by $\zeta\simeq1.0a$.

If we return to fig.~\ref{fig:correlation} where both the membrane and the
particle are moving, it is now understandable that an increasing diffusion
coefficient leads to a faster decrease of the correlation functions. An
analytic form for the observed correlation functions could not be found.

\begin{figure} 
  \begin{center}
    \includegraphics[width=0.9\linewidth,clip]{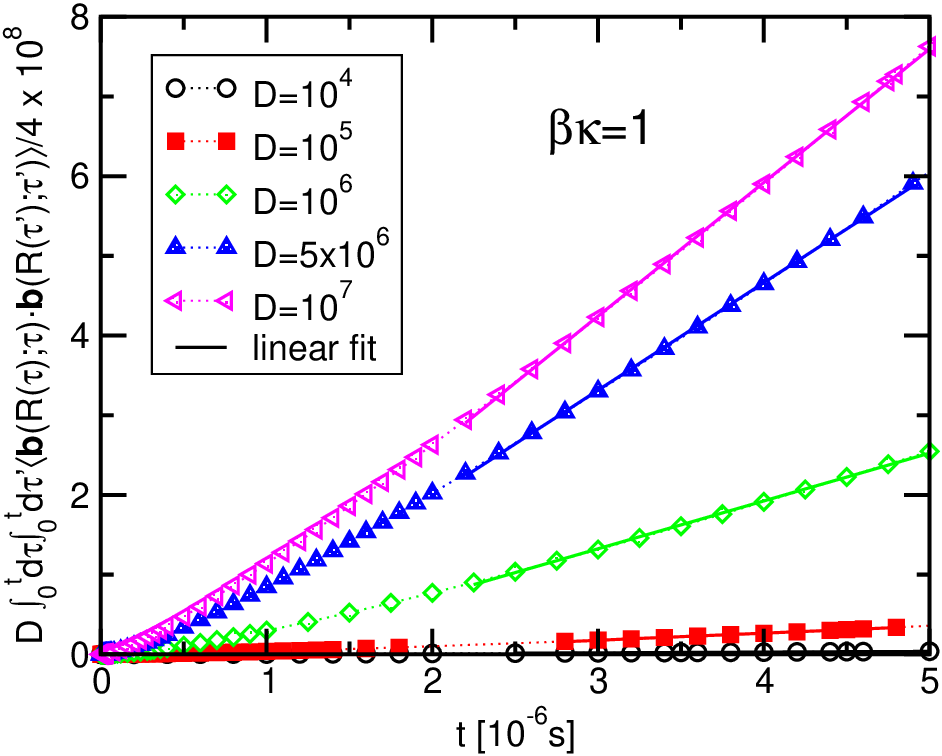}
  \end{center}
  \caption{\label{fig:msd_drift}(Color online) Additional mean square displacement
    $\langle\Delta\mathbf{R}^2_{\text{drift}}(t)\rangle/4D$ as a function of
    time $t$ for $\beta\kappa=1$ as derived by numerically evaluating the
    first term of eq.~\eqref{eq:msd_formal}.}
\end{figure}

In order to estimate the additional contribution to the projected diffusion
coefficient the correlation functions need to be integrated according to
eq.~\eqref{eq:msd_formal}. Regarding the results in fig.~\ref{fig:correlation}
it is obvious that the double integration of the correlation functions
$\langle b_{i}(\mathbf{R}(\tau);\tau)b_{i}(\mathbf{R}(\tau+\Delta
\tau);\tau+\Delta\tau)\rangle$ in time will cause a slower increase with
time for large $D$.  However, the additional contribution
$D_{\text{proj,drift}}/D$ is given by multiplying this integral by $D$. The
additional mean square displacement
$\langle\Delta\mathbf{R}^2_{\text{drift}}(t)\rangle/4D$ resulting from the
numerical integration of the results in fig.~\ref{fig:correlation} is
displayed in 
fig.~\ref{fig:msd_drift} for $\beta\kappa =1$.
For all regarded diffusion coefficients we find a linear behavior for large
times $t$.  Furthermore, the slope becomes larger for increasing $D$. In other
words the influence of the drift term becomes more pronounced the faster the
particle diffuses compared to the fluctuations of the membrane.

In this context it is interesting to check whether the additional contribution
is finite for an infinite intramembrane diffusion coefficient. For very large
diffusion coefficients the membrane will appear almost stiff for the moving
particle. This corresponds to the situation regarded in
fig.~\ref{fig:correlation_stiff}, where we found that the correlation function
is well described by an exponential function $\langle
b_{i}(\mathbf{R}(\tau);\tau)b_{i}(\mathbf{R}(\tau+\Delta
\tau);\tau+\Delta\tau)\rangle\simeq\langle
b_{i}^2(\mathbf{R}(\tau);\tau)\rangle\exp[-\sqrt{4D\Delta\tau}/\zeta]$. For
this particular function eq.~\eqref{eq:Dprojdrift_D} is easily evaluated. In
the limit of large times $t$ we find $D_{\text{proj,drift}}/D=\langle
b_{i}^2(\mathbf{R}(\tau);\tau)\rangle \zeta^2/2$. Thus
$D_{\text{proj,drift}}/D$ is independent of $D$, and only involves membrane
parameters. It therefore remains finite. For $\beta\kappa=1$ the largest
possible increase of $D_{\text{proj}}/D$ is approximately $\simeq0.015$.

\begin{figure}
  \begin{center}
    \includegraphics[width=0.9\linewidth,clip]{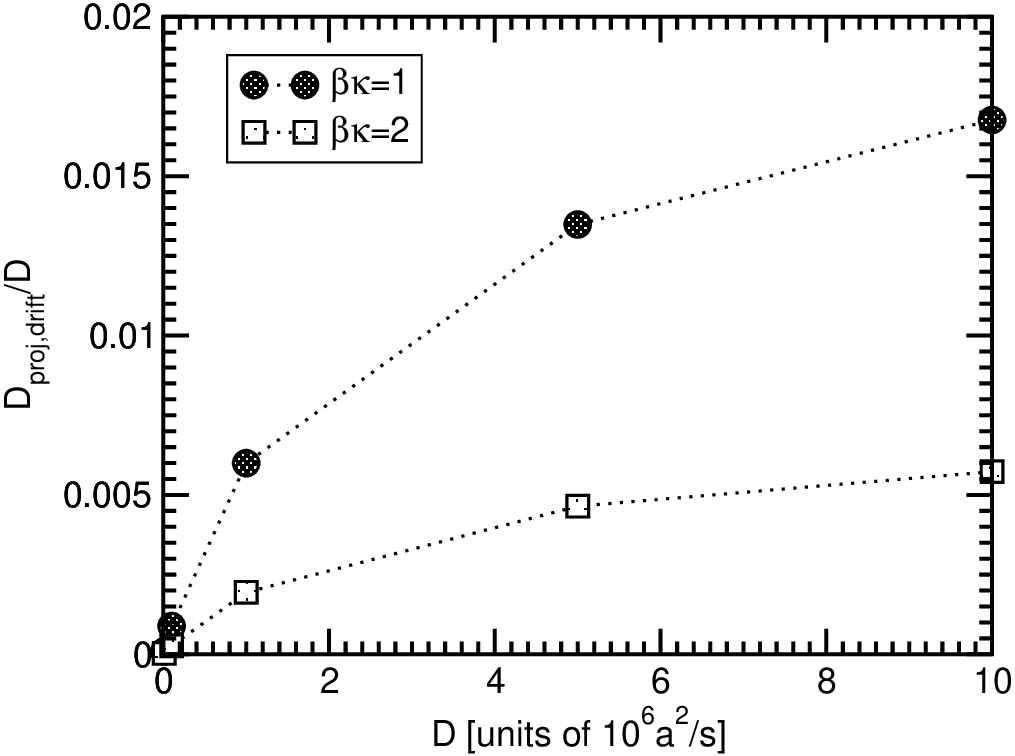}
  \end{center}
  \caption{\label{fig:bla}Additional contribution $D_{\text{proj,drift}}/D$
    following from linear fits to
    $\langle\Delta\mathbf{R}^2_{\text{drift}}(t)\rangle $, see
    fig.~\ref{fig:msd_drift}, as a function of $D$ for $\beta\kappa=1$,
    $2$.} 
\end{figure}

The additional terms to the ratio of projected to intramembrane diffusion
coefficient $D_{\text{proj,drift}}/D$ resulting from fits to
$\langle\Delta\mathbf{R}^2_{\text{drift}}(t)\rangle $, see
fig.~\ref{fig:msd_drift}, are displayed in fig.~\ref{fig:bla}.
Although an increase in $D$ causes a larger additional term we still see that
for the regarded diffusion coefficients that are much larger than those
experimentally observed in experiments, the additional term is always more
than two  orders of magnitudes smaller than $D_{\text{proj,preav}}/D$. For
$D=10^7a^2$/s the numerically determined value of $D_{\text{proj,drift}}/D$
agrees reasonably well with the previous estimate for infinite $D$.
Surprisingly, for situations when one expects the preaveraging
approximation to break down, it still gives reliable results. Even for
infinite diffusion coefficients the results for the projected diffusion
coefficient calculated within the preaveraging
approximation only differ from the actual values by less than two percent for
experimentally accessible membranes.  

After this discussion we can understand why our
simulation runs agree well with the preaveraging result for all $D$, as can be
seen in fig.~\ref{fig:Dproj_D_allD}. While our
simulation scheme is able to achieve an accuracy of a few percent
within reasonable computing time, the corrections caused by the drift term
cannot be identified directly via the mean square displacement of the
particles. Nevertheless, the explicit evolution of the membrane shape and
particle position make it possible to estimate this correction via the
evaluation of correlation functions.
\section{Conclusions}
In this paper we introduce a novel scheme that allows for the simultaneous
simulation of membrane fluctuations and intramembrane diffusion. The dynamics
of the system is expressed via the equation of motion for a membrane described
in the Monge gauge and the Langevin equation for a particle diffusing along a
surface whose form is given in the Monge gauge. The simulation algorithm
consists of the numerical integration of these two coupled differential
equations. To validate the membrane fluctuations we compare the height-height
correlation function determined from the simulations with known analytical
results. After ensuring that membrane fluctuation are reproduced correctly we
study free membrane bound diffusion along the membrane.  Since diffusion
coefficients are experimentally often determined from the projected path a
particle covers, we regard the ratio of the measured, projected diffusion
coefficient and the intramembrane diffusion coefficient that is a parameter of
the simulations and calculations. Both the simulations and the previous
calculations that apply a preaveraging approximation, show that the difference
between the measured and the true intramembrane diffusion coefficient is
largest for small bending rigidities $\kappa$ and small effective surface
tensions $\sigma$. This can be understood because small $\kappa$ and $\sigma$
lead to stronger membrane fluctuations.  Thus the actual path and the
projected path differ most.  Our calculations reveal a maximum reduction of
the projected diffusion coefficient by approximately 20 percent. We are aware
that the experimental corroboration of our findings is currently challenging,
but with the constantly increasing accuracy of methods to determine lateral
diffusion it should become feasible in the near future. Such experiments will
be important in showing that lateral diffusion is not only a function of the
direct interaction of lipids and proteins but also depends on material
properties of the membrane.

We also consider simulation runs with different intramembrane diffusion
coefficients $D$.  The subsequent analysis reveals a surprising observation:
the resulting ratios $D_{\text{proj}}/D$ all coincide independently of $D$.
Furthermore, the simulation results agree well with the analytical
preaveraging calculations. Only simulation runs with the smallest regarded
diffusion coefficients $D$ are expected to be well described by the analytical
results. For the largest used $D$, however, when diffusive and membrane time
scales become comparable, one would a priori assume that the ratio
$D_{\text{proj}}/D$ from the explicit simulations would differ from the
calculations.

An analysis of the Langevin equation that determines the movement of the
particle, demonstrates that correlations of the drift term caused by the
metric of the membrane are responsible for a possible increase in the measured
diffusion coefficient. In order to understand the applicability of the
preaveraging approximation we study these correlation functions using our
simulation scheme.  The relevant correlations decrease in time not only due to
membrane fluctuations but also due to the movement of the particle.  Our
simulations reveal that the influence of the drift term on the projected
diffusion coefficient increases with increasing $D$. But surprisingly for all,
even infinite, intramembrane diffusion coefficients it is very weak in
experimentally accessible membranes. In fact the influence is so small -- in
our simulations below two percent --, that it cannot be directly identified
from studying the mean square displacement of the particles within our scheme.
Only the study of the correlation functions of the drift term using our
simulations gives insight into the additional contributions to the projected
diffusion coefficient. For future studies one may now argue that preaveraging
suffices and the simulation scheme becomes unnecessary. Note that this
reasoning only makes sense as long as the particle diffuses freely. If an
additional interaction between membrane and protein is included the analytical
calculation of both the altered diffusion coefficient and membrane spectrum
relies on approximations which the simulations are capable of overcoming.

The experimental study of lateral diffusion in membranes has become a very
important and much noticed field that has revealed many different diffusion
phenomena. But so far only little complementing theoretical or simulational
work has been performed in order to develop a broader understanding of
experimental findings.  In previous work~\cite{Reister:2005} we regarded the
influence of a simple interaction between a diffusing particle and the
membrane curvature and found a strongly altered dependence of the projected
diffusion coefficient on bending rigidity and effective tension. It is
therefore promising that the measurement of lateral diffusion coefficients as
a function of membrane properties might shed light on the interaction between
protein and membrane.  While our previous calculations employed several
approximations the method introduced in this paper overcomes these.  Our
powerful simulation scheme is, therefore, a good starting point to investigate
the influence of various membrane-protein interactions not only on lateral
diffusion but also on the spectrum of the membrane.  Further effects that are
easily incorporated into our scheme are external influences on the membrane,
like tethers that resemble the attachment of a membrane to the cytoskeleton.
These and other extensions will be part of our future work.
\begin{acknowledgments} 
  The authors thank R.~Finken, T.~Speck, and O.~Farago for helpful discussions.
\end{acknowledgments} 


\end{document}